# Low-damping epsilon-near-zero slabs: nonlinear and nonlocal optical properties

Domenico de Ceglia,[1*] Salvatore Campione,[2] Maria Antonietta Vincenti,[1] Filippo Capolino,[2] and Michael Scalora[3]

[1]*National Research Council – AMRDEC, Charles M. Bowden Research Laboratory, Redstone Arsenal, AL, 35898, USA*

[2]*Department of Electrical Engineering and Computer Science, University of California Irvine, CA, 92697, USA*

[3]*Charles M. Bowden Research Laboratory, AMRDEC, US Army RDECOM, Redstone Arsenal, AL, 35898, USA*

[*]*domenico.deceglia@us.army.mil*

## ABSTRACT

We investigate second harmonic generation, low-threshold multistability, all-optical switching, and inherently nonlocal effects due to the free-electron gas pressure in an epsilon-near-zero (ENZ) metamaterial slab made of cylindrical, plasmonic nanoshells illuminated by TM-polarized light. Damping compensation in the ENZ frequency region, achieved by using gain medium inside the shells' dielectric cores, enhances the nonlinear properties. Reflection is inhibited and the electric field component normal to the slab interface is enhanced near the effective pseudo-Brewster angle, where the effective $\varepsilon \approx 0$ condition triggers a non-resonant, impedance-matching phenomenon. We show that the slab displays a strong effective, spatial nonlocality associated with leaky modes that are mediated by the compensation of damping. The presence of these leaky modes then induces further spectral and angular conditions where the local fields are enhanced, thus opening new windows of opportunity for the enhancement of nonlinear optical processes.

**I. INTRODUCTION**

Recent interest in ENZ materials has been motivated by the possibility of controlling antenna directivity[1,2] and achieving perfect couplers through electromagnetic tunneling in subwavelength, low permittivity regions.[3,4] ENZ materials may also be used to achieve enhanced harmonic generation,[5,6] optical bistability,[7,8] and soliton excitation .[9] The efficiency of harmonic generation is boosted in subwavelength ENZ slabs because the electric field is enhanced at the interface with a higher-index substrate.[5,6] This non-resonant enhancement occurs at oblique incidence for transverse magnetic (TM) polarization and is triggered by the continuity of the component of the displacement field normal to the interface.[10] At low irradiance levels, subwavelength ENZ slabs exhibit anomalous multistability and directional hysteresis,[7] so that even a weak nonlinearity can dominate the optical response. Although ENZ conditions occur naturally near the plasma frequency of any material (at visible and UV wavelengths for metals and semiconductors, and in the infrared range for dielectrics), artificial materials are advantageous because ENZ conditions may be engineered at virtually any wavelength. Waveguides operating near their cutoff frequency[11,12] and "rodded" media[11,13] exhibit plasma-like behavior by displaying near-zero effective permittivity. The strong electric dipole resonance of composite materials made of either periodic or random arrangements of metallic nanoparticles leads to ENZ effective condition.[14] The imaginary part of the permittivity limits the performance of ENZ materials for both linear and nonlinear optical applications. To circumvent this hurdle one may include gain material in the mixture, as suggested in Refs. 15 and 16. Active, fluorescent dyes introduced inside the cores of plasmonic nanoshells arranged in 3D periodic arrays may indeed suppress the imaginary part of the effective permittivity.[17] A mixture of metal-coated quantum dots dispersed in a dielectric matrix leads to similar behavior.[18-20] If the zero-crossing frequency of the real part of the effective permittivity turns out to be close to the center emission frequency of the active medium, then real and imaginary parts of the permittivity are simultaneously minimized. In this paper we exploit this approach to study an ENZ metamaterial based on a two-dimensional (2D) array of metallic, cylindrical nanoshells with gain medium embedded inside the cores. The array is designed by using effective medium approximation techniques, complex Bloch-mode analysis, and full-wave numerical simulations (summarized in Appendix A). The standard, local, homogenization procedures of the type discussed in Refs. 17 and 18 are used *only* as instrumental tools for the initial design of the bulk metamaterial properties because homogenization techniques adequately describe finite-thickness, metamaterial slabs of the kind described in this paper only for very small angles of incidence (5º or less). Strong spatial dispersion and fine spectral features sets in for larger angles, requiring a full-wave approach. This *effective* nonlocal behavior is strictly related to the *mesoscopic* nature of the array and its finite thickness, which is in turn enhanced by the very low damping in the system. Spatial dispersion phenomena are investigated by performing a complex Bloch-mode analysis of the finite-thickness array. In our approach we limited our study to the plane of incidence perpendicular to the plasmonic cylinders' axis, so that the electric dipole resonance becomes comparable to that achieved in three-dimensional (3D) arrays of spherical nanoshells.[21] For this reason transmittance, reflectance and absorption spectra that we report are similar to those observed in in Ref. 21, for 3D arrays of spherical nanoshells. However, in Ref. 21 the nature of the resonant spectral features is not analyzed or discussed. In contrast, here we expand the discussion by providing a full explanation of the physics behind these features and show how they influence nonlinear and nonlocal phenomena. In addition to a pseudo-Brewster mode responsible for the ENZ behavior of the slab, here we report additional low-damping, leaky modes that impact the slab response at oblique incidence. We thus identify two new channels for the enhancement of local fields and nonlinear processes in the low-damping regime: the first is related to the forced excitation of what we call a "pseudo-Brewster mode" near the pseudo-Brewster angle (Sec. II); the second is associated with the forced excitation of leaky modes supported by low-loss, finite-thickness ENZ slabs (Sec. III). We then turn to the discussion of nonlinear effects, i.e., enhancement of second harmonic generation (SHG) conversion efficiency originating from the metallic shells, low-threshold optical multistability and switching (Sec. IV). Even though we show that favorable conditions for SHG conversion efficiency are met near the pseudo-Brewster angle where local fields are maximized because of a forced excitation of the pseudo-Brewster mode, interestingly, we predict that optimal conditions for SHG conversion efficiency are rather met when *effective* nonlocal effects induced by leaky modes dominate the slab response. We then show that the large, angular and frequency selectivity of the tunneling effect triggered by the pseudo-Brewster mode, as well as the boost of electric field under these circumstances, lead to favorable conditions for enhancing self-phase modulation phenomena and inducing low-irradiance switching.

Finally, we discuss the role of *inherent* nonlocal effects induced by the free-electron gas pressure in the metallic nanoshells (Sec. V). The phenomenology of nonlocal contributions of free electrons on the optical response of nanoscale plasmonic structures has been widely discussed in literature.[22-28] Typical manifestations of the nonlocal, free-electron gas pressure are blue shift and broadening of plasmonic resonances, anomalous absorption,[29] unusual resonances above the plasma frequency,[25] and limitation of field enhancements.[28] These effects are more pronounced when the electron wavelength (~1 nm) becomes comparable to the radius of curvature of metallic nanostructures or to the distance between the metal boundaries of larger structures. Here we use both analytical and full-wave tools to show that these phenomena are magnified in ENZ arrays of metallic nanoshells, in the low-damping regime. We find that additional damping and limitations on field enhancements due to the inherent nonlocality arising from free-electron gas pressure may be mitigated by slightly increasing the gain in the nanoshells' cores.

## II. LINEAR PROPERTIES OF 2D ARRAYS OF CYLINDRICAL NANOSHELLS: FIELD ENHANCEMENT AT THE PSEUDO BREWSTER ANGLE

The geometry of a 2D array of cylindrical nanoshells is sketched in Fig. 1. Each cylinder has a core of radius $r_1$ with absolute permittivity $\varepsilon_1$, and a shell of external radius $r_2$ with absolute permittivity $\varepsilon_2$.

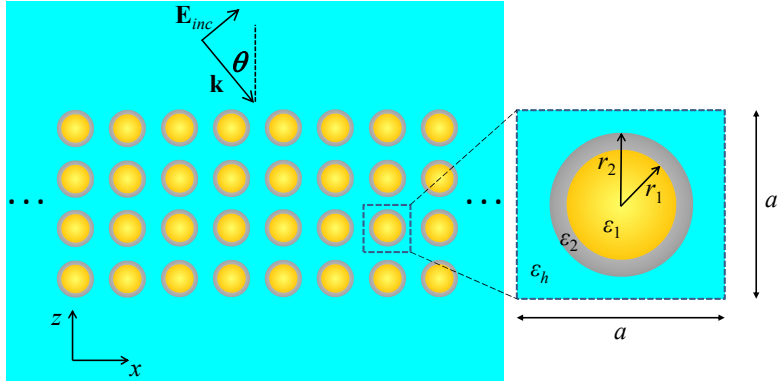

FIG. 1. Sketch of the ENZ slab illuminated by a TM wave. The slab is 4$a$ thick in the $z$ direction, and periodic in the $x$ direction with period $a$. The structure is assumed to be invariant in the $y$ direction. The unit cell (inset) consists of a cylinder with metallic nanoshell. The core of the shell is a mixture of a dielectric and an active, gain medium.

The source field is a plane wave with transverse magnetic (TM) field, i.e., electric field lying in the plane of incidence (x-z plane). The scattering problem is studied on the x-z plane (with period $a$ in both $x$ and $z$ directions); the structure is assumed to be invariant in the $y$ direction and surrounded by a medium with absolute permittivity $\varepsilon_h$. We retrieve the effective permittivity of the structure by using four methods, described in Appendix A: (1) Maxwell Garnett (MG) mixing formula; (2) quasi-static (QS) approximation; (3) complex Bloch mode analysis (MA); (4) Nicolson-Ross-Weir (NRW) retrieval method, based on full-wave numerical simulations and the inversion of the Fresnel formulas for transmission and reflections coefficients. The slab is composed of four layers of arrayed cylindrical nanoshells, is infinitely long in the $x$ direction, with thickness $d = 4a$ in the $z$ direction. The host material is a dispersion-free, silica-like medium with $\varepsilon_h = 2.25\ \varepsilon_0$. The frequency-dependent permittivity $\varepsilon_2$ of silver shell is taken from Ref. 30. For the purpose of lowering the metamaterial attenuation constant (i.e., lowering the loss coefficient), we assume the silica cores host Rhodamine 800 fluorescent molecules and that a pump signal alters the optical properties of the nanoshell cores. The volumetric dipolar excitation of the core is described via a four-level energy system as in Ref. 17, where the formula for permittivity is reported [see Eq. (14) in Ref. 17]. Following the notation in Ref. 17, we assume the following parameters: $\tau_{21} = 500$ ps, $\tau_{10} = \tau_{32} = 100$ fs, line width $\Delta\omega_a = 2\pi \times 15.9$ THz, central emission angular frequency $\omega_a = 2\pi \times 422$ THz, coupling constant $\sigma_a = 1.71 \times 10^{-7}$ C$^2$/Kg, volumetric dye concentration $\bar{N}_0 = 6.75$ mM, and pump rate $\Gamma_{pump} = 6.5 \times 10^9$ s$^{-1}$. The shell is $r_2 - r_1 = 5$ nm thick, and core radius is $r_1 = 25$ nm. The periodicity $a = 114$ nm is chosen so as the zero-crossing frequency of the real part of the effective permittivity matches the emission frequency of the gain material (422 THz). In Fig. 2 we plot the relative effective permittivity as a function of frequency evaluated by using the four homogenization methods described in Appendix A. All methods reveal a strong electric dipole resonance near 350 THz, and predict an ENZ permittivity near 422 THz. Molecular concentration and pump rate are chosen to balance Joule heating losses in the metal and gain in the active cores.

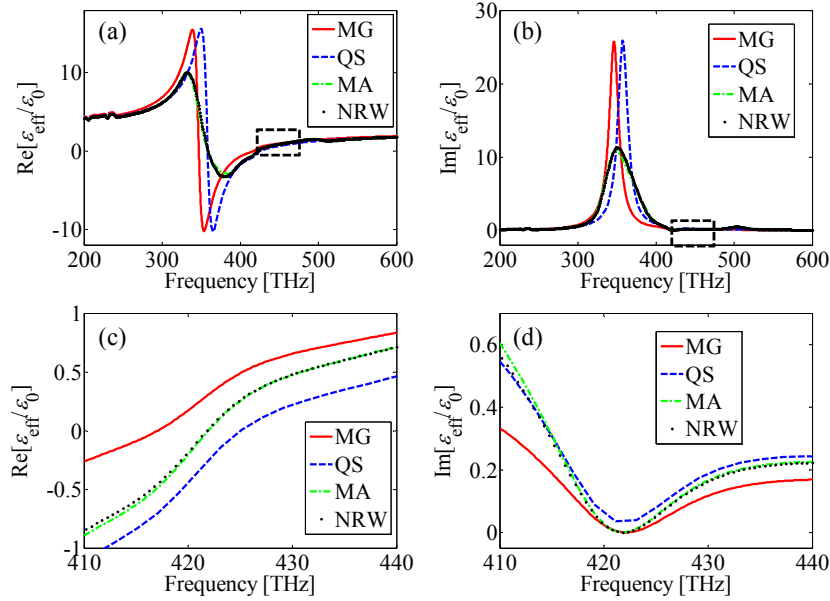

FIG. 2. (a) Real and (b) imaginary parts of the effective relative permittivities of the metamaterial sketched in Fig. 1 retrieved via the methods described in Appendix A. (c) and (d) are enlarged views of (a) and (b), respectively, near the zero-crossing point[dashed boxes in (a) and (b)]. MG=Maxwell-Garnett; QS=Quasi Static; MA=Mode Analysis; NRW=Nicolson-Ross-Weir.

We note in Fig. 2 that the complex Bloch-mode analysis and the NRW retrieval procedure yield almost identical effective parameters. These methods are considered the most accurate to characterize wave propagation since they are based on full-wave expansions. There are some intrinsic limitations in the MG and QS approximations, related to the fact that the period is not especially subwavelength: the MG mixing formula takes into account only dipolar contributions of the nanoshells and the effect of all mutual couplings is approximated, whereas the QS approximation neglects retardation effects and proper field distribution inside each elementary cell. In all cases we find that the complex $\varepsilon_{\text{eff}}$ approaches zero, as the active material compensates damping induced by the collective plasmonic resonance of the array. Damping compensation impacts slightly the real part of the effective permittivity, which would have a zero-crossing point even with higher damping. However, the active material reduces the imaginary part of the effective permittivity within the narrow band around 422 THz, as shown in Fig. 2(d), where the Lorentzian resonance of the active medium peaks. In Fig. 3 we report transmittance [Figs. 3(a) and 3(b)] and reflectance [Figs. 3(c) and 3(d)], defined as ratios of intensities, and absorptance [Figs. 3(e) and 3(f)] versus frequency and angle of incidence of a TM plane wave. Fig. 3 shows generally good agreement between analytical results for a finite metamaterial slab with homogenized permittivity obtained from the NRW procedure [Fig. 3(a), Fig. 3(c), Fig. 3(e)] and full-wave simulations obtained with the finite element method (FEM) [Fig. 3(b), Fig. 3(d), Fig. 3(f)]. These results prove that homogeneous medium modeling may be adopted to establish the linear properties of the slab in a relatively wide frequency range. However, in Fig. 3 we also observe additional features in the full-wave results near 422 THz, but we postpone their discussion to Sec. III. Two opposite features are observed and discussed here. The large impedance mismatch at the slab interface around 422 THz is due to the ENZ condition that increases the impedance near the emission frequency. In contrast, reflectivity is quenched within a very narrow angular bandwidth centered at the PB angle[31-33] relative to the interface between the surrounding silica-like medium and the slab. We refer to this region of low reflectance as the PB region. At the interface between a lossless medium with permittivity $\varepsilon_h$ and an ideal material slab with lossless permittivity $\varepsilon_{slab}$, reflected power for TM waves is minimized at the Brewster angle, $\theta_B = \tan^{-1}\left(\sqrt{\varepsilon_{slab}/\varepsilon_h}\right)$. However, absorption losses prevent the reflection from vanishing, and a minimum reflection angle known as PB angle[31-33] exists, where the impedance matching condition between the two media is approached. The analytical expression for the PB angle at the interface between a lossless medium with index $n_h = \sqrt{\varepsilon_h / \varepsilon_0}$ and a lossy material with complex index $n_{slab} + ik_{slab} = \sqrt{\varepsilon_{slab} / \varepsilon_0}$ is found in Refs. 31-33:

$$\theta_{PB} = \cot^{-1}\left\{\left[\sqrt{\xi}\left(\cos\zeta + \sqrt{3}\sin\zeta\right) - \frac{1}{3}\right]^{1/2}\right\}, \qquad (1)$$

with $\zeta = \frac{1}{3}\cos^{-1}\left(\frac{\psi}{\xi^{3/2}}\right)$, $\xi = \left(\frac{n_h^2}{n_{slab}^2 + k_{slab}^2}\right)^2 + \frac{1}{9}$, $\psi = \left(\frac{n_h^2}{n_{slab}^2 + k_{slab}^2}\right)^3 \left(\frac{n_{slab}^2 - k_{slab}^2}{n_{slab}^2 + k_{slab}^2}\right) + \frac{1}{27}$. In Fig. 4(a) we plot the contours of constant PB angle with $\varepsilon_{slab}/\varepsilon_0$ varying in the complex plane, assuming $\varepsilon_h/\varepsilon_0 = 2.25$. The figure shows that when the slab displays ENZ response the PB angle approaches normal incidence. At the PB angle one expects inhibition of reflection and increased transmission and absorption. The effect is captured in Fig. 3 as a narrow bright spot in transmission and absorption maps, and as a narrow dark spot in the reflection map in the near-zero permittivity region. Frequency and angle selectivities of this phenomenon are strictly related to the value of Im[$\varepsilon_{slab}/\varepsilon_0$] and to the steepness of Re[$\varepsilon_{slab}/\varepsilon_0$] near the zero-crossing frequency. In Fig. 4(b) we report an enlarged view of the transmission map in Fig. 3(a), based on the homogenized slab approximation (with parameters retrieved via the NRW method in Fig. 2), and we overlap the frequency-dependent PB angle calculated using Eq. (1). The PB angle curve follows the maximum transmission angle even for a four-layers-thick slab, adding credence to the idea that non-resonant, quasi-impedance-matching occurs at the interface between the transparent substrate and the ENZ slab at near-zero input angles.

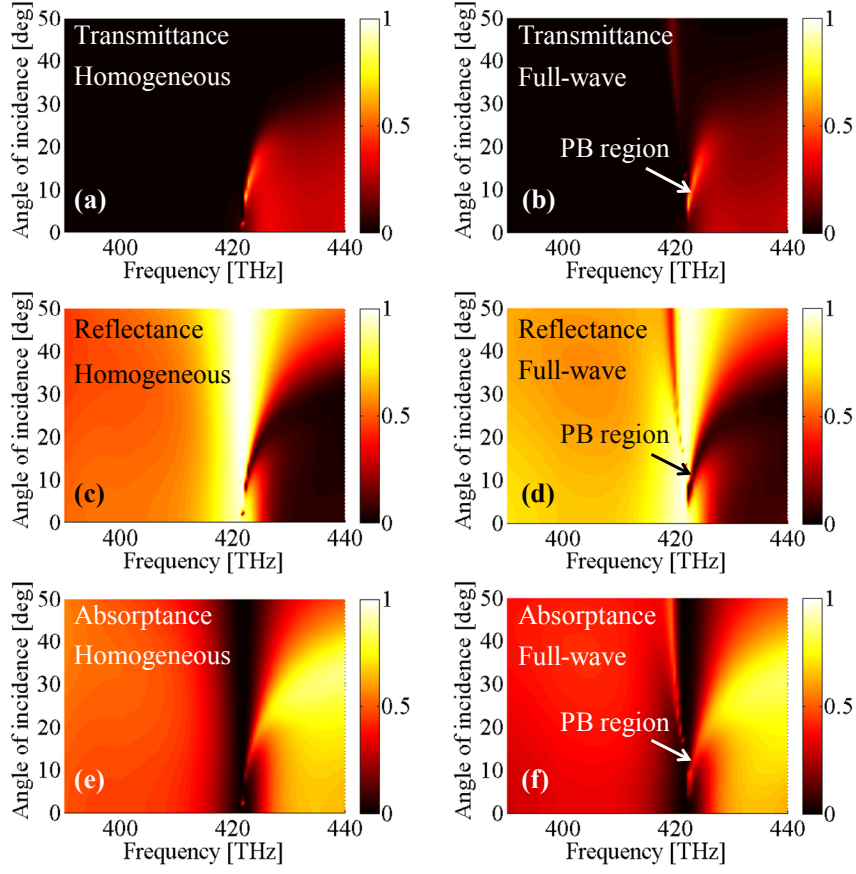

FIG. 3. Transmittance [(a), (b)], reflectance [(c), (d))] and absorptance [(e), (f)] through the slab of Fig. 1 versus frequency and angle of incidence calculated by: (i) NRW homogenization procedure, and (ii) using full-wave numerical simulations (FEM). The PB region centered at the PB angle is indicated with an arrow.

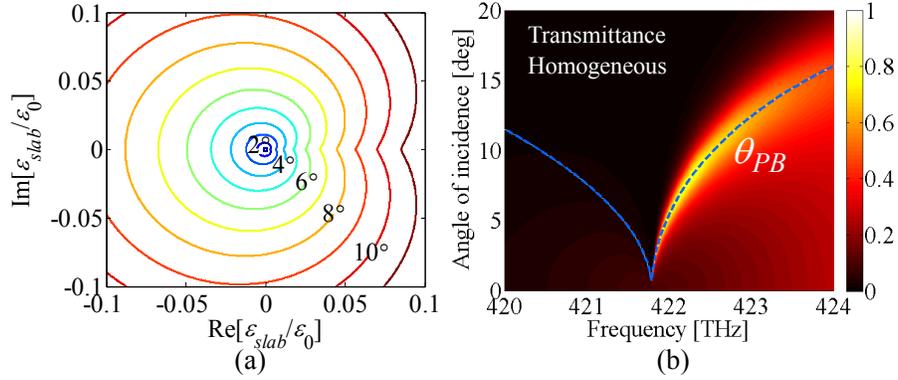

FIG. 4. (a) Pseudo Brewster angle ($\theta_{PB}$) contours with $\varepsilon_{slab}$ varying in the complex permittivity plane. (b) Transmission through a homogeneous metamaterial slab with $\varepsilon_{slab}$ taken from the NRW retrieval method. The dashed blue line is the PB angle curve $\theta_{PB}$ evaluated with Eq. (1).

As theoretically demonstrated in Ref. 10, a singular field enhancement in finite-thickness ENZ slabs may be obtained for critical angle condition, for total transmission condition (Brewster angle) and in the limit of simultaneously vanishing values of effective permittivity and angle of incidence. Plane-wave excitation of the ENZ slab in Fig. 1 at the PB angle [Eq. (1)] defines a real angle condition, even for lossy slabs, very close to the total transmission and the critical angle conditions defined in Ref. 10 by means of complex incident angles. For this reason, in the following we will only refer to the real PB angle condition and show how it leads to enhanced electric fields and nonlinear phenomena (Sec. IV).

**III. EFFECTIVE NONLOCALITY: ADDITIONAL MODES OF THE ENZ SLAB**

In the previous section we showed that the dominant spectral effect in the ENZ slab of Fig. 1 is a tunneling phenomenon at oblique incidence for TM polarization, in the PB region. The panels in Fig. 5 are enlarged views of the maps in Fig. 3, obtained by using higher frequency and angular resolutions to allow narrow spectral features to emerge. The discrepancy between results from full-wave and homogenization methods in Fig. 5 cannot be appreciated in Fig. 3. The differences are due to narrow resonances mediated by additional *forced* modes[34] in the low-damping spectral region around 422 THz, i.e., within the emission bandwidth of the gain medium. These novel *forced* modes are excited around the PB region and generate Fano-like spectral features,[35] observable either as narrow asymmetric transmission (or reflection) profiles, or selective enhanced absorption regions (Fig. 5). We stress that these additional spectral features are not predicted and are absent in the homogenized slab model described in Fig. 5(a) and 5(c).

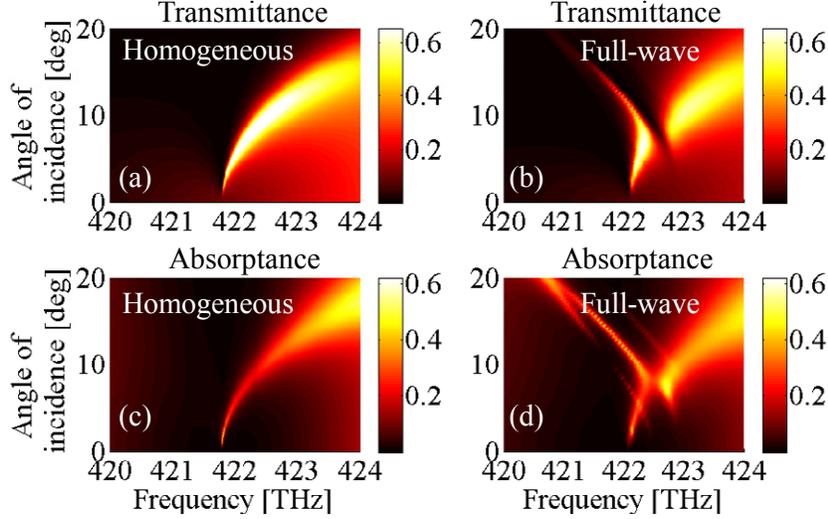

FIG. 5. Transmittance [(a), (b)] and absorptance [(c), (d)] through the slab in Fig. 1 versus frequency and angle of incidence calculated via NRW homogenization and full-wave numerical simulations as noted.

The nature of these new spectral features is revealed by investigating the *free*, complex Bloch modes supported by the slab in the *x* direction. These modes are found by using the complex Bloch mode analysis technique described in Appendix A. We find the modes with Bloch wave vector in the direction $\hat{\mathbf{v}} = \hat{\mathbf{x}}$ and set a unit cell with lattice translation vector $\mathbf{R} = \hat{\mathbf{x}} a$ as in Fig. 6(a). The slab is sandwiched by the silica-like host medium as in Fig. 1, and the unit cell is terminated in the $\pm z$ directions with perfectly matched layers[36] adapted to the host medium. These terminations "absorb" leaky modes, i.e., radiating modes within the light cone of the host medium with real part of the complex transverse wavenumber $|\beta_x| < k_h$ (where $k_h$ is the host wavenumber). These modes can affect transmission and absorption via phase-matching with the plane wave incident from the host medium, as described in Fig. 1. A *free* mode is perfectly matched to an external field when both real and imaginary parts of the transverse wavenumber $k_x^{\text{mode}} = \beta_x + i\alpha_x$ are matched to the complex wavenumber of the incident field. The mode cannot be excited by a simple homogeneous plane wave (a source is necessary to excite a *free* mode [34]). However, an incident plane wave may *force the excitation* of a mode by phase matching with the real part $\beta_x$ of the complex wavenumber $k_x^{\mathrm{mode}}$. This can modify transmission and absorption properties when the imaginary part $\alpha_x$ of the modal wavenumber is small. In Figs. 6(b) and (c) we report the dispersion curves of real and imaginary parts of the Bloch's wave number of *free* modes with low imaginary part. In this region the slab supports four leaky modes ($|\beta_x|/k_h < 1$) whose $\alpha_x$ nearly vanish, also thanks to low damping conditions. The dispersion of the mode labeled PB is close to the dispersion of the dominant mode in the unbounded metamaterial described in Sec. II. This mode plays a role in the PB tunneling phenomenon, as shown in Fig. 4. We refer to the other three supported modes as M1, M2 and M3. Given the very small imaginary part of their wavenumbers at frequencies smaller than 422 THz, modes M1-M3 are the prevailing response of the slab in this spectral region.

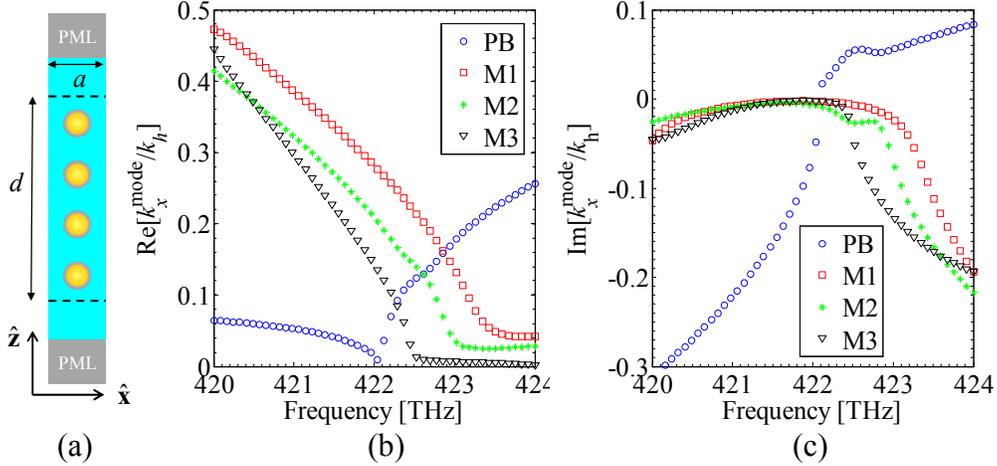

FIG. 6. (a) Unit cell used for the complex Bloch MA of the finite-thickness ENZ slab in Fig. 1. The slab is periodic along *x* with period *a*, and finite in the *z* direction (height *d* = 4*a*). (b) Real and (c) imaginary part of the wavenumber of the complex modes supported by the slab normalized to the host medium wavenumber $k_h$.

The phase matching angle for each of these modes, i.e., the angle at which the transverse wavenumber of the impinging plane wave matches the phase constant of the modes, is calculated as $\theta^{\mathrm{mode}} = \sin^{-1}(\beta_x/k_h)$. The angle-frequency dispersion curve associated with these modes [white stars in Fig. 7(a)] overlaps well with the narrow resonances visible in the frequency-angle absorptance map of the ENZ slab, as shown in Fig. 7(a). We note that the analytical expression of the PB angle in Eq. (1) predicts the tunneling angle through the slab only in the limit of the effective medium approximation shown in Fig. 4(b). Instead, the PB mode dispersion in Fig. 7(a), evaluated via Bloch theory for the unit cell in Fig. 6(a), is in excellent agreement with the more complicated dispersion of the PB tunneling (transmission) region resulting from full-wave numerical simulations. In general, the impact of the free modes on the spectral response of the ENZ slab is significant when the imaginary part $\alpha_x$ is low. This is further enhanced when the effective material damping is lowered thanks to the active material in the nanoshells' cores. However, it should be pointed out that both the low-damping ENZ condition and the excitation of the leaky modes M1, M2 and M3 may be similarly attained by using any low-loss plasmonic material, without resorting to active gain inclusions within the metamaterial.

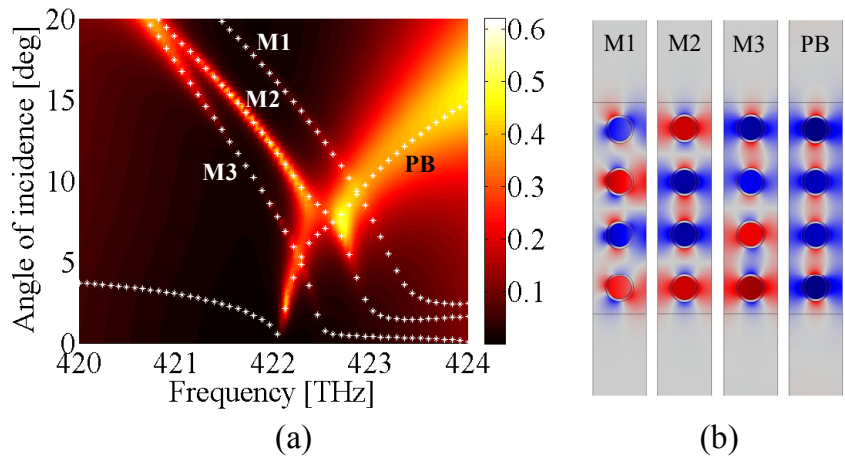

FIG. 7. (a) Absorptance map of the ENZ slab evaluated via FEM. The dispersion of the modes (white stars) is superimposed, showing the correlation with absorptance. (b) Real part of the electric field $z$-component Re[$E_z$] for the four leaky modes supported by the slab. The field distribution for free modes M1, M2 and M3 is taken at 421.5 THz, while the field distribution for the PB mode is taken at 423.5 THz. The color scale (arbitrary units) in each panel of (b) is adapted to highlight the maxima (deep red) and the minima (deep blue) of Re[$E_z$].

For example, semiconductor-based oxides have been indicated as intrinsic low-loss plasmonic materials for the near-infrared frequency range.[37] Overviews of alternative plasmonic materials based on doped metals, doped semiconductors, metal alloys and band-structure engineering are presented in Refs. 38 and 39.

The presence of the modes M1-M3 and the PB mode thus makes it difficult to homogenize the slab with a simple, local effective medium approach, either the NRW approximation or the complex Bloch MA for the unbounded metamaterial (see Sec. A.3 and Sec. A.4 in Appendix A) is adequate for angles up to 5° (Fig. 5), where the dominant mode of the slab is the PB mode. For larger angles one should take into account the finite nature of the slab's thickness and perform a complete modal analysis of the slab as described above. An effective nonlocal model used to homogenize the ENZ slab may account for these additional modes (modes M1-M3), in addition to the PB mode, and their dispersion. For example, in Ref. 40 an effective nonlocal model of an anisotropic ENZ slab composed of nanorods was derived by fitting full-wave numerical simulations to account for spatial dispersion effects and the presence of additional waves. In Fig. 7(b) we report the distribution of the real part of the phasor of the electric field component $E_z$ for the four modes supported by the slab. The fields for modes M1-M3 are plotted at 421.5 THz, whereas the field of the PB mode is plotted at 423.5 THz. The PB mode is characterized by an alignment of the dipoles magnitude and phase along the $z$ direction. Similarly, the electric dipoles are oriented along the $z$ direction even for the additional three modes. However, these modes experience a phase difference between elements in the $z$ direction. Indeed, some dipoles are oriented along + $z$ (blue color) and others along - $z$ (red color). The parity symmetry of the additional modes is essential to determine the degree of interaction with the PB mode. While modes M1 and M3 display weak interaction with the PB mode, strong coupling between modes M2 and PB leads to the anti-crossing behavior at $\approx 422.5$ THz and for an angle of incidence near 7°. A simple coupled-mode theory[41] argument may be used to explain this phenomenon. Modes

M1 and M3 display odd symmetry with respect to the center of the slab, and couple weakly to the even-parity, tunneling field of the PB mode. The opposite occurs for the interaction between M2 and the PB mode: both show even symmetry. We stress that the number of additional modes, their spectral positions and dispersions strictly depend on slab thickness, i.e., the number of periods in the $z$ direction. Under these circumstances, the definition and the determination of an effective nonlocal permittivity are not straightforward, and may not even be necessary. Homogenization including nonlocal effects is discussed in Refs. 42 and 43, but certain structures also require the introduction of Drude type transition layers or sheets that account for the transition between free space and the homogenized bulk metamaterial,[44] further complicating the analysis. For this reason we perform a complex Bloch mode analysis of the finite-thickness structure that provides a full physical interpretation of the spectra retrieved via full-wave numerical simulations, as shown in Fig. 7. Moreover, the excitation of additional modes by phase-matched plane waves is inhibited when damping in the metamaterial is large since modal attenuation constants would be larger. A more detailed discussion of the nature and dispersion of these additional leaky modes is not the focus of this work and hence postponed to future investigations.

**IV. ENHANCEMENT OF HARMONIC GENERATION AND LOW-THRESHOLD OPTICAL BISTABILITY**

The efficiency of nonlinear processes generally depends on local field intensity. In subwavelength structures the requirement of high field localization is essential. We now show the amount of field magnification that can occur in subwavelength ENZ slabs. In previous sections we have demonstrated that two different linear phenomena may be observed in a slab composed of cylindrical nanoshells under low-damping conditions: (1) an impedance-matching,

PB effect, and (2) the presence of narrow resonances mediated by additional slab modes. In both cases the slab displays enormous field enhancement accompanied by strong field localization around the nanoshells, as shown in Fig. 7(b). The efficiency of harmonic generation is thus expected to increase significantly. As described in Refs. 10 and 45, homogeneous ENZ slabs with small $\text{Im}(\varepsilon_{eff})$ display transmittance spectra with narrower angular selectivity. In addition we find that the strong chromatic dispersion of the PB mode wavenumber relative to the 2D array of nanoshells (see Fig. 2) leads also to a pronounced frequency-selective transmission process. In Fig. 8 we plot the angle-frequency map of the maximum electric field enhancement factor (EF) for the same slab of cylindrical nanoshells with active gain analyzed in Sec. II and Sec. III.

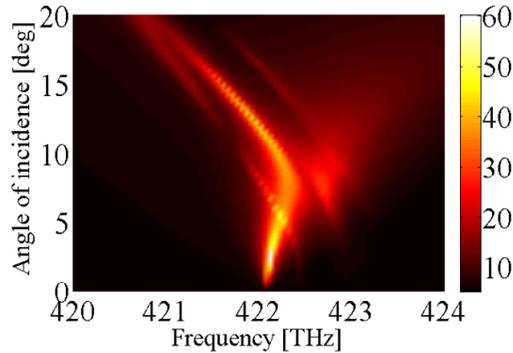

FIG. 8. Maximum electric field enhancement factor (EF) inside the slab shown in Fig. 1 versus frequency and angle of incidence around the zero crossing point of the real part of the effective permittivity.

The maximum EF is defined as $\text{EF} = \max\left[\left|\mathbf{E}_{ENZ}(\mathbf{r})\right|\right]/E_0$, where $\left|\mathbf{E}_{ENZ}(\mathbf{r})\right|$ is the electric field amplitude inside the ENZ slab (evaluated by means of full-wave, FEM simulations), and $E_0$ is the amplitude of the incident plane wave. We have also verified that a plot of the EF, where EF is now averaged over each elementary cell, reproduces the same trend shown in Fig. 8, after replacing the scale maximum with 22 instead of 60. We observe that the peaks of maximum EF follow the dispersion of the PB mode and the three additional modes in Fig. 6. For the dye

concentration and pump rate used in this paper one has a remarkable $\text{EF} \approx 60$, mainly due to $z$-polarized fields. For comparison, we stress that $\text{EF} \approx 1$ in the absence of gain, in the same range.

## A. Boosting SHG from the plasmonic nanoshells at the PB angle

By adopting the hydrodynamic model for free electrons in the silver nanoshells, the induced current density obeys the following spatio-temporal differential equation[46]

$$\frac{\partial \tilde{\mathbf{J}}_f}{\partial t} = -\gamma_f \tilde{\mathbf{J}}_f + \frac{\tilde{n}_f e^2}{m^*}\tilde{\mathbf{E}} + \frac{1}{\tilde{n}_f e}\left[\tilde{\mathbf{J}}_f \nabla \cdot \tilde{\mathbf{J}}_f + \left(\tilde{\mathbf{J}}_f \cdot \nabla\right)\tilde{\mathbf{J}}_f\right] - \frac{\mu_0 e}{m^*}\tilde{\mathbf{J}}_f \times \tilde{\mathbf{H}} + \frac{e}{m^*}\nabla \tilde{p}, \qquad (2)$$

supplemented by the continuity equation $\partial \tilde{n}_f / \partial t = -e\nabla \cdot \tilde{\mathbf{J}}_f$. Here $\tilde{\mathbf{J}}_f$ is the instantaneous free electron current density; $\gamma_f = 7.284 \times 10^{13}$ s$^{-1}$ is the damping coefficient of free electrons;[47] $e$ is the electron charge; $m^* = m_e = 9.109 \times 10^{-31}$ Kg is the effective electron mass; $\mu_0$ is the vacuum magnetic permeability; $\tilde{n}_f$ is the instantaneous free electron density ($n_0 = 4.963 \times 10^{22}$ cm$^{-3}$ being the equilibrium free electron density[47]); $\tilde{\mathbf{H}}$ and $\tilde{\mathbf{E}}$ are the magnetic and electric fields, respectively; and $\tilde{p}$ is the electron gas pressure. Here the tilde on top of variables denotes time domain quantities. The spatial dependence of the current and electron densities, the fields, and gas pressure is implicit in Eq. (2). The first two terms on the right hand side of Eq. (2) account for the linear Drude response of conduction electrons; the third term describes convective forces acting on free electrons; the fourth term is the magnetic Lorentz force; the last term is due to gas pressure. SHG is studied at low input irradiances of the fundamental field, and so we adopt the undepleted pump approximation. We postpone the analysis of nonlocal contributions to harmonic generation to Sec. V.

We transform Eq. (2) to the frequency domain by expressing all variables as a superposition of two components, oscillating at the fundamental (FF) and second harmonic (SH) frequencies, as in Refs. 21, 48 and 49. Volume and surface current contributions at the SH frequency are expressed as functions of the optical properties of silver and the dielectric in contact with the metal surfaces. The linear response at the SH frequency due to free and bound electrons is extracted from the experimental silver permittivity.[30] We first find a solution for the electromagnetic problem at the FF using the FEM. We then use the FF field to evaluate the current sources for the SH field. The problem is solved again with the FEM applied to a slab that contains four nanocylinders in the $z$ direction, and by limiting the computational space to one period in the $x$ direction [Fig. 6(a)] by setting Floquet boundary conditions. Plane-wave matched ports are used at $z = \pm 500$ nm ($z = 0$ is the center of the slab). The irradiance of the FF input plane wave is $I_{in} = \sqrt{\varepsilon_h / \mu_0} E_0^2 / 2 = 100$ kW/cm$^2$. SHG efficiency is defined as

$$\eta_{SH} = \frac{\left| \int_{-a/2}^{a/2} S_z(\omega_{SH}) \big|_{FW} dx \right| + \left| \int_{-a/2}^{a/2} S_z(\omega_{SH}) \big|_{BW} dx \right|}{a I_{in} \cos\theta}, \qquad (3)$$

where $S_z(\omega_{SH})\big|_{FW}$ and $S_z(\omega_{SH})\big|_{BW}$ are the $z$-components of the SH, time-averaged Poynting vectors in the forward (transmission) and backward (reflection) regions, respectively, evaluated just below and above the slab. Hence, it represents the total power generated at the SH frequency, regardless of the direction of propagation. In Fig. 9(a) we plot the SHG efficiency $\eta_{SH}$ on a logarithmic color scale as a function of the FF and angle of incidence. As in the EF map, the SHG efficiency closely follows the dispersion of the modes of the slab, i.e., the PB mode and the additional M1-M3 modes. The efficiency in this angle-frequency domain is relatively large, given the weak nature of the bulk and surface quadratic, metal nonlinearities involved, and the relatively modest FF input irradiance. The absolute maximum of SHG efficiency is $\approx 10^{-8}$ at an incidence angle of $\approx 12°$, which may be associated to leaky mode M2. For comparison, this is well over six orders of magnitudes larger than conversion efficiency obtained for flat metal surfaces illuminated by TM-polarized light. This result thus emphasizes the importance of the leaky modes M1-M3 discussed in Sec. III for enhanced harmonic generation, for which we observe the largest efficiency. In order to understand the role that damping compensation plays in the harmonic generation process, it is instructive to compare SHG efficiency for two slabs, with and without fluorescent dye molecules inside the cores. We do so in Fig. 9(b), where we report $\eta_{SH}$ as a function of the incident angle in the range 0-30° for two different scenarios that involve a FF tuned at 422.3 THz: (i) the structure in Fig. 1 and the parameters described in Sec. II, including the active gain from the fluorescent dye molecules that leads to very low damping; (ii) the same structure without gain in the core, i.e., assuming the cores' permittivity of $\varepsilon_1 = \varepsilon_h$, and therefore damping is due to losses in the metal. The result in Fig. 9(b) shows that low-damping conditions boost SHG efficiency by nearly four orders of magnitudes relative to the same structure without gain.

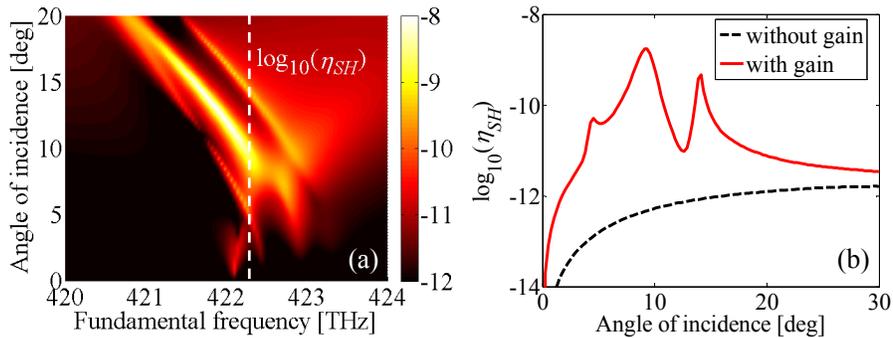

FIG. 9. (a) Angle-frequency SHG efficiency map for the ENZ slab with gain. (b) SHG efficiency with and without gain. The FF is 422.3 THz. The solid red curve is a projection of the efficiency map along the white dashed line in (a) (FF=422.3 THz). The dashed black curve is obtained in absence of gain, assuming $\varepsilon_1 = \varepsilon_h$ in the shells' core.

By combining this result with the one in Fig. 9(a) one may infer that this harmonic generation enhancement is due to the forced excitation of the PB and the additional modes supported by the ENZ slab.

**B. Optical multistability at the PB angle**

For angles of incidence smaller than $\approx 5°$ and frequencies in the range 420-424 THz, the main response of the ENZ slab is due to the PB mode, as one may infer by inspecting either Fig. 5 or Fig. 7(a). In this small angular region the homogenization of the slab with an effective medium approach is valid (Sec. II), since additional modes M1-M3 do not significantly affect the incident plane wave. In order to investigate multistability near the PB angle we consider the slab described by the homogeneous effective permittivity reported in Fig. 2, obtained via the NRW retrieval method. We assume the slab exhibits an effective, cubic, nonlinear susceptibility $\chi_{\text{eff}}^{(3)} = 10^{-17} \text{ m}^2/\text{V}^2$, compatible, for example, with the nonlinear susceptibility of silver composites and doped polymers.[50] The constitutive relation then includes self-phase modulation, so that the electric displacement field in the slab is:

$$\mathbf{D} = \varepsilon_{\text{eff}} \mathbf{E} + \varepsilon_0 \chi_{\text{eff}}^{(3)} |\mathbf{E}|^2 \mathbf{E}. \qquad (4)$$

The nonlinear polarization is represented by the second term on the right hand side of Eq. (4). There are two concurrent conditions that trigger low-threshold multistability: (i) the nonlinear polarization is amplified near the PB angle thanks to local field intensity enhancement; (ii) the nonlinear polarization may easily become the dominant response of the ENZ slab, as the linear permittivity vanishes.[7, 51] Multi-valued solutions of transmittance, reflectance and absorptance at the fundamental frequency are retrieved by using the graphical method outlined in Ref. 52. We consider a fixed, small incidence angle ($\theta = 2.9°$), and incident frequencies in a region (420.5-423 THz) that includes the zero crossing point of $\text{Re}(\varepsilon_{eff})$.

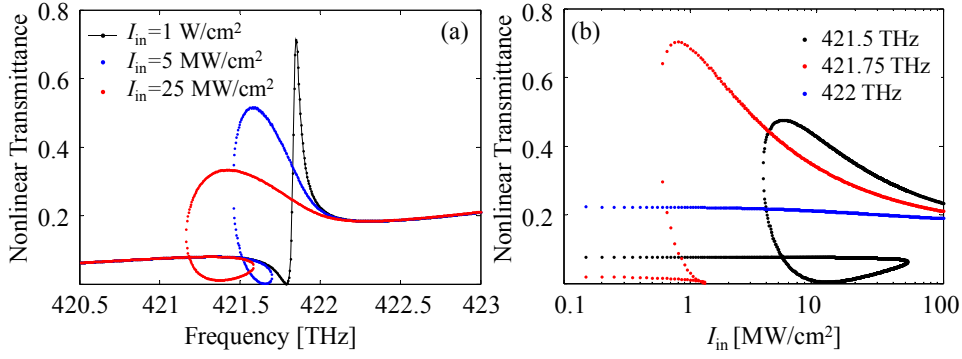

FIG. 10. (a) Multistable transmission spectra through the ENZ slab for three input irradiances. (b) Transmittance as a function of input irradiance for three different values of input frequency. A low-threshold switching ($I_{in} \approx 1$ MW/cm$^2$) is obtained at 421.75 THz.

The nonlinear transmittance, defined as the ratio of intensities, is plotted in Fig. 10(a) for three different irradiance levels $I_{in}$. While the transmittance for $I_{in}$ = 1 W/cm$^2$ is mono-stable and nearly overlaps the linear transmittance curve (not shown here for brevity), when $I_{in}$ = 5 MW/cm$^2$ the spectrum shows a multistable region between 421.5 THz and 421.75 THz that widens further when $I_{in}$ = 25 MW/cm$^2$. The results in Fig. 10(a) show that the input irradiance may be used as a control parameter to shift the PB angle of the slab, which in turn may be employed as a low-power optical switch. This is illustrated in Fig. 10(b). The angle of incidence is set at $\theta = 2.9°$, $I_{in}$ varies from 0.1 MW/cm$^2$ to 100 MW/cm$^2$, and we consider three input frequencies. At 421.5 THz the slab presents a wide hysteresis loop and the switching threshold is $\approx 10$ MW/cm$^2$. For 421.75 THz the hysteresis loop is narrower and nonlinear transmittance switches from a low state (<1%) to a high state (nearly 60%) at a much lower irradiance threshold ($\approx 1$ MW/cm$^2$). At 422 THz we do not observe hysteresis or switching. We thus surmise that in the neighborhood of the PB angle multistable behavior of the ENZ slab is very sensitive due to near-zero values of the linear effective permittivity and the sharp selectivity of the slab in both frequency and angular domains.

## V. IMPACT OF THE INHERENT NONLOCAL RESPONSE OF THE NANOSHELLS

The ENZ metamaterial slab outlined above was optimized to boost field enhancement and nonlinear phenomena. In particular, the amount of damping compensation regulated by the volumetric dye concentration was tailored to minimize losses based on a local response for the 5-nm-thick metallic nanoshells. In this section we return to the linear properties of the slab and include *inherent* nonlocal effects. All nonlinear terms present in the hydrodynamic model, Eq. (2), are neglected while the electron gas pressure contribution $(e/m^*)\nabla\tilde{p}$ is retained. If one treats the free electron plasma as a Thomas-Fermi gas, one may then link $\tilde{p}$ to the macroscopic polarization and to the free electron current density. The nonlocal differential equation for the free electron current density is[53]

$$\frac{3}{5}v_{\text{Fermi}}^2 \nabla(\nabla \cdot \mathbf{J}_f) + (\omega^2 + i\omega\gamma_f)\mathbf{J}_f = \frac{i\omega n_0 e^2}{m^*}\mathbf{E}. \qquad (5)$$

We assume the Fermi velocity of free electrons in silver is $v_{\text{Fermi}} = 1.39 \times 10^6$ m/s. Equation (5) is coupled to the standard Helmholtz equation for the electric field which accounts for the response of bound electrons.[53] As an additional boundary condition for the current density, required by the presence of the nonlocal term in Eq. (5), we impose $\hat{\mathbf{n}} \cdot \mathbf{J}_f = 0$ at the two concentric circular boundaries limiting the metallic shells (see Fig. 1), where $\hat{\mathbf{n}}$ is the unit vector normal to these boundaries. The effective relative permittivity is then evaluated with the FEM, using the NRW technique (see Sec. A.4 in Appendix A) with the addition of nonlocal effects as in Eq. (5). The results are shown in Fig. 11(a) in the frequency range that includes the $\text{Re}(\varepsilon_{\text{eff}})$ zero-crossing point. For comparison we also report the relative effective permittivity (calculated via NRW) reported in Fig. 2 in the limit of the local theory for free electrons, i.e., assuming $v_{\text{Fermi}} = 0$ in Eq. (5). We have tested the consistency of these results against a nonlocal version of MG effective medium approximation that includes the hydrodynamic pressure term in the metallic nanoshells and the additional boundary condition $\hat{\mathbf{n}} \cdot \mathbf{J}_f = 0$ in the expression of the first order electric scattering coefficient, as described in Ref. 54 (see Appendix A). The results are shown in Fig. 11(b), where both the local [same as red line in Fig. 2(a)] and nonlocal, MG effective permittivities are plotted. The predictions of the MG approximation are in good qualitative agreement with those based on the NRW technique. The main discrepancy between the two models, i.e., a shift of the zero-crossing frequency of $\text{Re}(\varepsilon_{\text{eff}})$, is due to the fact that the MG approximation neglects multipolar contributions and mutual coupling effects while the NRW method is based on a full-wave expansion of the fields, as discussed in Sec. II. Regardless of the intrinsic differences between the two homogenization techniques, we can reach similar conclusions on the nonlocal effects due to the free-electron gas pressure: we observe a slight frequency blue shift ($\approx 0.5$ THz) of the zero-crossing point and a significant variation of $\text{Im}(\varepsilon_{\text{eff}} / \varepsilon_0)$ from $\sim 10^{-4}$ to $\sim 10^{-2}$. Both phenomena may be explained by analyzing the nonlocal behavior of a single cylindrical nanoshell. For example, a blue-shift of the nonlocal scattering cross section of coated spheres was demonstrated in Ref. 22 using a semi-classical, infinite-barrier model. A similar blue-shift and near-field quenching of plasmons in dimers and thin metallic waveguides was predicted in Ref. 23.

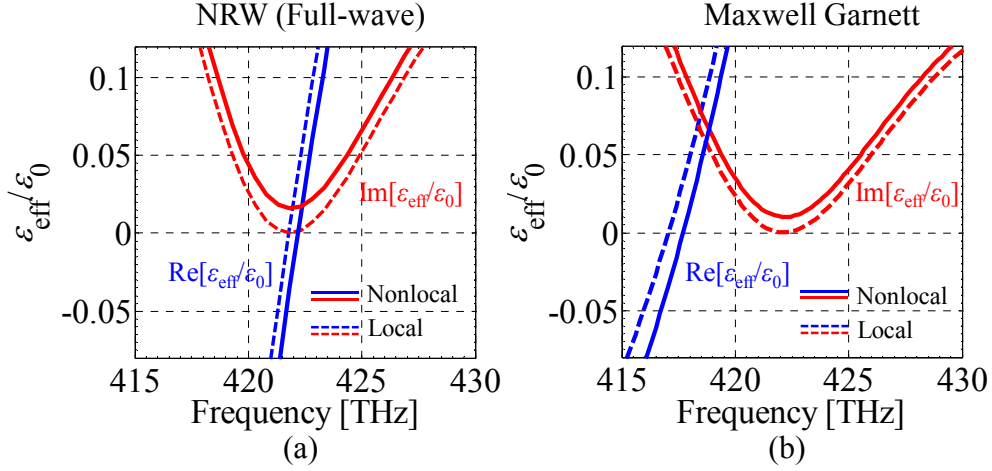

FIG. 11. (a) Local (dashed lines) and nonlocal (solid lines) effective parameters of the ENZ slab retrieved with the NRW technique. Dynamics of free electrons in the nonlocal model is treated by using Eq. (5). (b) Local (dashed lines) and nonlocal (solid lines) effective parameters of the ENZ slab retrieved with the Maxwell Garnett effective medium approximation.

Analogous descriptions of *inherent* nonlocal phenomena in core-shell nanoparticles can be found in Refs. 24, 54-56. We ascribe the differences reported in Fig. 11(a) and 11(b) to the nonlocal blue shift of the cylindrical-shell dipole resonance. The weak, nonlocal perturbation acting on the single plasmonic resonator is magnified in the array. To evaluate the macroscopic impact of the nonlocal corrections on the PB effect of the slab we compare the angle-frequency, transmittance maps by homogenizing the ENZ slab using both local and nonlocal NRW models. The result is shown in Figs. 12(a) and (b). While the performance of the ENZ slab in terms of field enhancement and impedance matching suffers little as the band blue-shifts, the nonlocality increases effective damping and hinders the beneficial effects of the active gain in the shell's core. This is inferred from Fig. 12(a) and (b), by observing the differences between the transmission maps evaluated with and without nonlocal contributions. The detrimental effects of nonlocality manifest themselves as a decreased transmittance near the PB angle, and broadening of transmittance spectra in both frequency and angular domains.

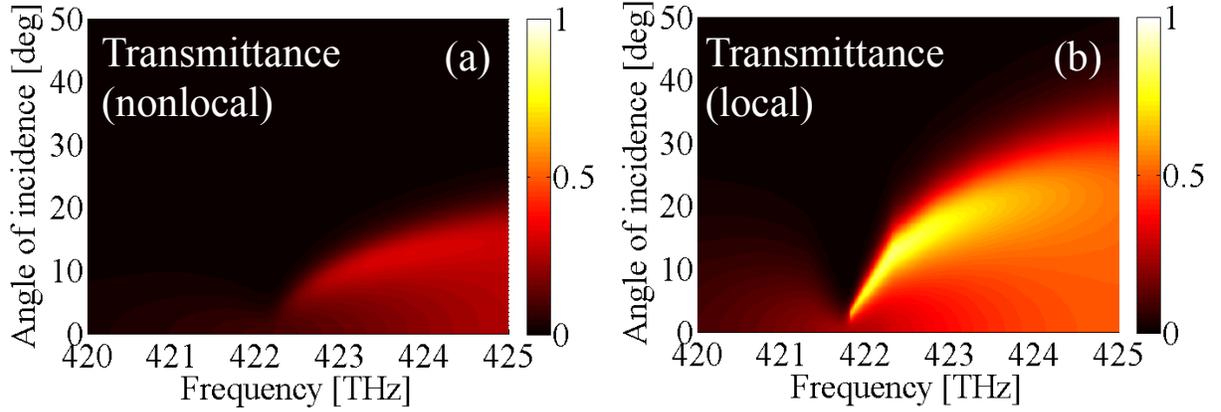

FIG.12. Transmission maps versus frequency and angle of incidence, using nonlocal (a) and local (b) models. The slab is homogenized with the NRW technique including (a) and neglecting (b) the nonlocal term $\frac{3}{5}v_{\text{Fermi}}^2 \nabla(\nabla \cdot \mathbf{J}_f)$ in Eq. (5).

The narrow resonances mediated by leaky modes M1-M3 (see Fig. 7(a)) are also affected by the *inherent* nonlocality, as we have verified using full-wave numerical simulations that include the nonlocal model [Eq. (5)] for free electrons. In particular, we recognize two effects related to the increased damping due to the nonlocality: (i) spectral and angular broadening of the resonances mediated by the modes M1-M3, as well as the PB mode, and (ii) a decrease of the *maximum* EF from $\approx 60$ in the limit of local response for free electrons to $\approx 15$ when the nonlocality is included. This implies a corresponding decrease of the *average* EF from $\approx 22$ to $\approx 4$, which directly impacts nonlinear phenomena by lowering their efficiency. However, by adopting a simplistic but qualitative description one may note that this detrimental effect is due to an increase of the imaginary part of the permittivity, and in the small detuning of the real part of the permittivity, so that the magnitude of $\varepsilon_{\text{eff}}/\varepsilon_0$ is slightly larger. Nevertheless, one may still exploit the main features of the inclusion of gain in the shells' cores to rebalance the negative effect of nonlocality. For example, in this particular case a slight increase of dye concentration from 6.75 mM to 6.975 mM neutralizes the extra damping due to electron gas pressure, thus helping to restore both the effective low-damping ENZ and correspondingly strong electric field EF to values we predicted in the limit of the local model.

## VI. CONCLUSIONS

We have investigated the field enhancement capabilities of an ENZ slab illuminated at oblique incidence with TM-polarized light. The metamaterial is composed of a 2D array of cylindrical nanoshells. In particular, we consider ENZ materials with low-damping that can be obtained by resorting to active gain material included in the core of the shells. Low-damping (i) favors impedance matching and electric field enhancement near the pseudo-Brewster angle of the slab and (ii) triggers an effective nonlocal response mediated by additional leaky modes observable as narrow, resonant, Fano-like states in the angular-frequency transmission or absorption spectra. We demonstrated that these low-damping induced effects may be exploited to boost second harmonic generation from metallic nanoshells. We also showed low-threshold optical multistability and switching by exploiting large field enhancement of the non-resonant, pseudo-Brewster mode for small angles of incidence. Finally, we investigated the role of the nonlocal response associated with free-electron gas pressure and, while we observed lowering of field enhancement factors near the zero-crossing point, we also found that the use of an active gain material can compensate additional damping due to the nonlocality.

## ACKNOWLEDGMENTS

This research was performed while the authors M. A. Vincenti and D. de Ceglia held a National Research Council Research Associateship award at the U. S. Army Aviation and Missile Research Development and Engineering Center.

# APPENDIX A. RETRIEVAL METHODS FOR THE EFFECTIVE PERMITTIVITY OF THE METAMATERIAL SLAB

The complex effective permittivity of the structure in Fig. 1 has been retrieved by using the four different methods detailed below. Comparisons are also provided and discussed in the main body of the paper.

## A.1 Maxwell Garnett (MG) effective medium theory

We calculate the polarizability of a single cylindrical nanoshell by assuming only dipolar contributions. For subwavelength radii one may replace the nanoshell with an isotropic line-dipole, whose induced dipole moment (in the *x-z* plane) is expressed as

$$\mathbf{p} = \alpha_e \mathbf{E}_{loc} = \alpha_e \left( \mathbf{E}_{inc} + \mathbf{E}_s \right), \qquad (6)$$

where $\alpha_e$ is the nanoshell electric polarizability, assumed isotropic in the *x-z* plane, $\mathbf{E}_{loc}$ is the local field at the nanoshell location, given by the sum of the incident field $\mathbf{E}_{inc} = E_0 \left( \sin\theta \hat{\mathbf{x}} + \cos\theta \hat{\mathbf{z}} \right)$ and the one scattered by all the other cylindrical nanoshells ($\mathbf{E}_s$); $\theta$ is the angle of incidence, $E_0$ is the plane wave amplitude, and $\hat{\mathbf{x}}$, $\hat{\mathbf{z}}$ are unit vectors in the *x* and *z* directions, respectively. The fields have implicit time dependence $\exp(-i\omega t)$. In the dipole approximation, $\alpha_e = -i 8\varepsilon_h D_1 / k_h^2$ [57], where $k_h = k_0 \sqrt{\varepsilon_h / \varepsilon_0}$ is the wavenumber in the host medium, $k_0$ is the free-space wavenumber, $\varepsilon_0$ is the free-space absolute permittivity $\varepsilon_h$ is the absolute host permittivity, $D_1$ is the first order electric scattering coefficient. The coefficient $D_1$ is calculated using Mie theory by imposing matching boundary conditions on electric and magnetic fields tangential at each interface of the annular structure. The analytical expression of $D_1$ may be found in Refs. 57 and 58 within the approximation of the local theory for free electrons in the metallic shells. However, the nonlocal hydrodynamic pressure term and the additional boundary condition $\hat{\mathbf{n}} \cdot \mathbf{J}_f = 0$ on the nanoshell's boundaries can be easily included in the Mie theory, as described in Ref. 54, where the nonlocal corrections to the coefficient $D_1$ are given. Once the polarizability is known, the evaluation of the absolute effective permittivity of the 2D array follows by applying the MG mixing rule

$$\varepsilon_{\text{eff}} = \varepsilon_h \left[ 1 + \left( \frac{\varepsilon_h}{N} \left( \frac{1}{\alpha_e} + i \frac{k_h^2}{8\varepsilon_h} \right) - L \right)^{-1} \right], \tag{7}$$

where $N = 1/S$ is the number of cylinders per surface unit, $S$ is the area of the unit cell, and $L=1/2$ is the depolarization factor for the circular symmetry of the particle (coated cylindrical inclusion). Note that in Eq. (7) we subtract the radiation loss term $-ik_h^2/(8\varepsilon_h)$ from $1/\alpha_e$ (which includes radiation losses) as suggested in Ref. 59, so as to cancel out radiation damping for a two-dimensional array with periodicity smaller than $\lambda_0/2$.

### A.2 Quasi-static (QS) approximation

The QS retrieval method is described in Ref. 60 and here briefly summarized. An external, static electric field $\mathbf{E}_{DC}$ is applied to the unit cell of the array, i.e., to the square domain $[-a/2, a/2] \times [-a/2, a/2]$ in Fig. 1. The electric potential $\phi(\mathbf{r})$ is found by solving Poisson's equation $\nabla \cdot [\varepsilon_\omega(\mathbf{r}) \nabla \phi(\mathbf{r})] = 0$, where $\varepsilon_\omega(\mathbf{r})$ is the frequency-dependent absolute permittivity at position $\mathbf{r}$ in the cell. The $xx$ component of the effective permittivity tensor $\varepsilon_{\text{eff},xx}$ is retrieved by modeling the unit cell as a nanocapacitor and by applying a static voltage $V_x$ between the two virtual plates at $x = -a/2, a/2$. The boundary conditions at $z = -a/2, a/2$ are set as $E_{z,QS} = -\partial\phi/\partial z = 0$. The capacitance per unit length is given by $C_x = \varepsilon_{\text{eff},xx} = q_o/V_x$, where $q_o$, the charge per unit length on the plate $x = a/2$, is found by integrating the charge density $\sigma(z) = -\varepsilon_h \partial\phi(x,z)/\partial x|_{x=a/2}$ along $z$. Thus, the QS absolute effective permittivity is found by $\varepsilon_{\text{eff},xx} = \int_{-a/2}^{a/2} \sigma(z) dz / V_x$. The $zz$ component of the effective permittivity tensor $\varepsilon_{\text{eff},zz}$ may be evaluated by applying a capacitor potential $V_z$ along the $z$ direction and following the same procedure used for $\varepsilon_{\text{eff},xx}$; however the nanoshell symmetry imposes $\varepsilon_{\text{eff},xx} = \varepsilon_{\text{eff},zz} = \varepsilon_{\text{eff}}$.

**A.3 Complex Bloch mode analysis (MA)**

For TM polarization, the Helmholtz equation for the magnetic field reads as follows:

$$\nabla \cdot \left[\frac{1}{\varepsilon(\mathbf{r},\omega)} \nabla H_y(\mathbf{r})\right] + \omega^2 \mu(\mathbf{r},\omega) H_y(\mathbf{r}) = 0, \qquad (8)$$

where $\varepsilon(\mathbf{r},\omega) = \varepsilon(\mathbf{r}+\mathbf{R},\omega)$ and $\mu(\mathbf{r},\omega) = \mu(\mathbf{r}+\mathbf{R},\omega)$ are periodic functions with lattice translation vector $\mathbf{R}$. The unit cell of the metamaterial is illustrated in Fig. 1. By applying Bloch's theorem, solutions of Eq. (8) may be written as $H_y(\mathbf{r}) = u(\mathbf{r})\exp(-i\mathbf{k}_B \cdot \mathbf{r})$ where $\mathbf{k}_B$ is the Bloch wave vector restricted to the first Brillouin zone, and $u(\mathbf{r}) = u(\mathbf{r}+\mathbf{R})$ is a periodic function. The weighted residual expression of Eq. (8), obtained by using Galerkin's method,[61] is integrated over a closed domain and the divergence theorem is applied to write the weak formulation of Eq. (8) according to the FEM. Once the angular frequency is set and a particular Bloch wave vector direction $\hat{\mathbf{v}}$ fixed, a quadratic, matrix eigenvalue equation in $k_B = \mathbf{k}_B \cdot \hat{\mathbf{v}}$ is solved using COMSOL Multiphysics. Details about this procedure may be found in Refs. 62-64. In order to retrieve the effective parameters we first set $\hat{\mathbf{v}} = \hat{\mathbf{x}}$ and $\mathbf{R} = \hat{\mathbf{x}}a$ and find the dominant complex Bloch mode, i.e., the one whose complex $k_B$ has the smallest imaginary part (for more details see Ref. 65). The effective absolute permittivity corresponding to this mode is calculated as $\varepsilon_{\text{eff}} = \varepsilon_0 \left( k_{B,d} / k \right)^2$, where $k = \omega / c$ is the wavenumber *in vacuo*.

## A.4 Nicolson-Ross-Weir (NRW) retrieval method

Following Refs. 66 and 67, we consider a slab of finite thickness along *z* and an infinite number of periods in the *x* direction, as in Fig. 1. Transmission (*T*) and reflection (*R*) coefficients at normal incidence ($\theta = 0$) are calculated as functions of frequency using the FEM via COMSOL Multiphysics. The effective absolute permittivity of the slab is

$$\varepsilon_{\text{eff}} = \varepsilon_0 \left\{ \pm \frac{1}{kd} \left[ \cos^{-1}\left( \frac{1 - R^2 + T^2}{2T} \right) + 2\pi m \right] \right\}^2, \tag{9}$$

Where $m = 0, \pm 1, \pm 2, \ldots$, $d = 4a$ is the slab thickness. Equation (9) is obtained by considering a unitary relative permeability; *m* is determined by following the procedure described in Ref. 68.